\documentclass[fleqn,usenatbib]{mnras}

\usepackage[T1]{fontenc}

\DeclareRobustCommand{\VAN}[3]{#2}
\let\VANthebibliography\thebibliography
\def\thebibliography{\DeclareRobustCommand{\VAN}[3]{##3}\VANthebibliography}

\usepackage{graphicx}	
\usepackage{amsmath}	
\usepackage{amssymb}

\def\lcii{$L_{\rm [CII]}$}

\def\lir{$L_{\rm IR}$}

\def\zphot{\ifmmode z_{\rm phot}\else$z_{\rm phot}$\fi}

\def\Cii{[C\,{\sc ii}]}

\def\ltsima{$\buildrel<\over\sim$}
\def\la{\lower.5ex\hbox{\ltsima}~}
\def\gtsima{$\buildrel>\over\sim$}
\def\ga{\lower.5ex\hbox{\gtsima}~}
\def\deg~{$^{\circ}$}

\title[CII line and Dust constraints of GN-z11]{NOEMA observations of GN-z11: Constraining Neutral Interstellar Medium and Dust Formation in the Heart of Cosmic Reionization at $z=10.6$}

\author[Y. Fudamoto et al.]{
Y. Fudamoto,$^{1,2}$\thanks{E-mail: yoshinobu.fudamoto@gmail.com}
P. A. Oesch,$^{3,4}$
F. Walter,$^{5}$
R. Decarli,$^{6}$
C. L. Carilli,$^{7}$
A. Ferrara,$^{8}$
L. Barrufet,$^{3}$\newauthor
R. Bouwens,${^9}$
M. Dessauges-Zavadsky,$^{3}$
E. J. Nelson,$^{10}$
H. Dannerbauer,$^{11}$
G. Illingworth,$^{12}$
A. K. Inoue,$^{1,13}$\newauthor
R. Marques-Chaves,$^{3}$
I. P{\'e}rez-Fournon,$^{11,14}$
D. A. Riechers,$^{15}$
D. Schaerer,$^{3}$
R. Smit,$^{16}$
Y. Sugahara,$^{1,2}$\newauthor
P. van der Werf,$^{17}$
\\
$^{1}$Waseda Research Institute for Science and Engineering, Faculty of Science and Engineering, Waseda University, 3-4-1 Okubo, Shinjuku, Tokyo 169-8555, Japan\\
$^{2}$National Astronomical Observatory of Japan, 2-21-1, Osawa, Mitaka, Tokyo, Japan\\
$^{3}$Department of Astronomy, University of Geneva, Chemin Pegasi 51, 1290 Versoix, Switzerland\\
$^{4}$International Associate, Cosmic Dawn Center (DAWN)\\
$^{5}$Max Planck Institute for Astronomy, K{\"o}nigstuhl 17, 69117 Heidelberg, Germany\\
$^{6}$INAF – Osservatorio di Astrofisica e Scienza dello Spazio di Bologna, via Gobetti 93/3, I-40129, Bologna, Italy\\
$^{7}$National Radio Astronomy Observatory, P. O. Box 0,Socorro, NM 87801, USA\\
$^{8}$Scuola Normale Superiore, Piazza dei Cavalieri 7, 50126 Pisa, Italy\\
$^{9}$Leiden Observatory, Leiden University, NL-2300 RA Leiden, Netherlands\\
$^{10}$Department for Astrophysical and Planetary Science, University of Colorado, Boulder, CO 80309, USA\\
$^{11}$Instituto de Astrof{\'i}sica de Canarias (IAC), E-38205 La Laguna, Tenerife, Spain\\
$^{12}$Department of Astronomy and Astrophysics, University of California, Santa Cruz, CA 95064, USA\\
$^{13}$Department of Physics, School of Advanced Science and Engineering, Faculty of Science and Engineering, Waseda University, 3-4-1, Okubo, Shinjuku\\ , Tokyo 169-8555, Japan\\
$^{14}$Universidad de la Laguna, Dpto. Astrofísica, E-38206 La Laguna, Tenerife, Spain\\
$^{15}$I. Physikalisches Institut, Universit{\"a}t zu K{\"o}ln, Zülpicher Strasse 77, D50937 K{\"o}ln, Germany\\
$^{16}$Astrophysics Research Institute, Liverpool John Moores University, 146 Brownlow Hill, Liverpool L3 5RF, United Kingdom\\
$^{17}$Leiden Observatory, Leiden University, P.O. Box 9513, 2300 RA Leiden, The Netherlands\\
}

\date{Accepted XXX. Received YYY; in original form ZZZ}

\pubyear{2023}

\begin{document}
\label{firstpage}
\pagerange{\pageref{firstpage}--\pageref{lastpage}}
\maketitle

\begin{abstract}
We present results of dust continuum and \Cii$\,158\,{\rm \mu m}$ emission line observations of a remarkably UV-luminous ($M_{\rm UV}=-21.6$) galaxy at $z=10.603$: GN-z11. 
Using the Northern Extended Millimeter Array (NOEMA),  observations have been carried out over multiple observing cycles. We achieved a high sensitivity resulting in a $\lambda_{\rm rest}=160\,{\rm \mu m}$ continuum $1\,\sigma$ sensitivity of $13.0\,\rm{\mu Jy/beam}$ and a \Cii\ emission line $1\,\sigma$ sensitivity of $31\,\rm{mJy/beam\,km/s}$ using $50\,\rm{km/s}$ binning with a $\sim 2\,{\rm arcsec}$ synthesized beam.
Neither dust continuum nor \Cii$\,158\,{\rm \mu m}$ line emission are detected at the expected frequency of $\nu_{\rm [CII]} = 163.791\,\rm{GHz}$ and the sky location of GN-z11.
The upper limits show that GN-z11 is neither luminous in \lir\ nor $L_{\rm [CII]}$, with a dust mass $3\,\sigma$ limit of ${\rm log}(M_{\rm dust}/{\rm M_{\odot}}) < 6.5-6.9$ and with a \Cii\ based molecular gas mass $3\,\sigma$ limit of ${\rm log}(M_{\rm mol,[CII]}/{\rm M_{\odot}}) < 9.3$.
Together with radiative transfer calculations, we also investigated the possible cause of the dust poor nature of the GN-z11 showed by the blue color in the UV continuum of GN-z11 ($\beta_{\rm UV}=-2.4$), and found that $\gtrsim3\times$ deeper observations are crucial to study dust production at very high-redshift.
Nevertheless, our observations show the crucial role of deep mm/submm observations of very high redshift galaxies to constrain multiple phases in the interstellar medium.
\end{abstract}




\section{Introduction} \label{sec:intro}
The first few hundred million years after the Big Bang at redshifts of $z > 10$ are the last major unexplored epoch in the history of the Universe. 
Recently, the arrival of the JWST has opened a completely new window to observe the very high redshift Universe. JWST/NIRCam’s unprecedented sensitivity at infrared wavelengths has enabled confident photometric information on $z_{\rm phot} > 10$ candidates detected over multiple bands \citep[e.g.,][]{Castellano2022,Finkelstein2022,Naidu2022,Hainline2023,Harikane2023}.
At the same time, spectroscopy with JWST obtained multiple emission lines in the rest-frame UV to optical wavelength, providing unambiguous spectroscopic redshifts \citep[e.g.,][]{Roberts-Borsani2022,Morishita2022,Cameron2023,Bunker2023b,Boyett2023,Matthee2023,Williams2023} and unprecedented diagnostic power of interstellar medium (ISM) properties for the first time \citep[e.g.,][]{Carnall2023,Nakajima2023,Tacchella2022,Sanders2023,Laseter2023}.
This
started to reveal a detailed picture of galaxy formation at $z>10$, which was extremely challenging to reach before the JWST era \citep[see][for a review]{Madau2014}.

GN-z11 is the highest-redshift galaxy identified using the {\it Hubble Space Telescope (HST)} and the {\it Spitzer Space Telescope} \citep{Oesch2016}.
GN-z11, was first identified from {\it HST} imaging data from the CANDELS survey \citep[][]{Bouwens2010,Grogin2011,Koekemoer2011}.
In particular, it lies in the CANDELS DEEP area in the GOODS-North field at RA, Dec$ = $ 12:36:25.46,\,+62:14:31.4 (J2000). GN-z11 was found to be the brightest galaxy candidate at $z>10$ \citep{Bouwens2015,Oesch2015}.
From HST WFC3/IR G141 slitless grism spectroscopy, the galaxy had a redshift of $z_{\rm grism}=11.09^{+0.08}_{-0.12}$  through the identification of the Ly$\alpha$ continuum break \citep{Oesch2016}. 

Due to the extremely high UV-luminosity of $M_{\rm UV}=-21.6$, detailed JWST observations of GN-z11 have already been performed.
The JWST Advanced Deep Extragalactic Survey (JADES; \citealt{Eisenstein2023}) obtained deep JWST NIRSpec slit spectroscopy and confirmed its redshift to be $z=10.6034\pm0.0013$ based on detections of multiple emission lines \citep{Bunker2023}.
Together with photometric information obtained with JWST NIRCam imaging, studies confirmed that GN-z11 is the most UV-luminous galaxy at $z>10$ known to date \citep[e.g.,][]{Harikane2023,Hainline2023}.
In particular, the extremely compact morphology ($R_{\rm e}=64\pm20\,{\rm pc}$; \citealt{Tacchella2023}) and the large UV luminosity ($M_{\rm UV}=-21.6\pm0.02$; \citealt{Tacchella2023}) suggest that GN-z11 could represent very early onset of AGN activity \citep{Maiolino2023a} or that it shows possible signatures of population-III stars in its surrounding blob \citep{Maiolino2023b}. 
At the same time, the significant detection of nitrogen emission lines (NIV]\,1486\AA\ and NIII]\,1750\AA) and the large over-abundance seen in the N/O ratio could require a new scenario of nitrogen production and/or oxygen depletion \citep{Cameron2023}, such as the formation of supermassive stars through stellar collisions in an extremely dense environment \citep{Charbonnel2023}. 

With the newly obtained detailed observations from JWST/NIRCam and JWST/NIRSpec, GN-z11 remains one of the most important galaxies at $z>10$ to understand gas fueling mechanism for galaxy growths and dust production in the early Universe. 
To fully understand the formation of this remarkably luminous galaxy and to constrain the formation mechanisms of massive galaxies at $z>10$, it is essential to investigate multiple components of its interstellar medium (ISM) as well as dust formation and dust-obscured star formation activity. In addition to the rest-frame UV images and spectra, information about the dust and neutral ISM properties from rest-frame far-infrared (FIR) spectroscopy is thus required to provide a comprehensive picture arising from observations covering a wide wavelength range \citep[see][for reviews]{Carilli2013,Hodge2020}.


Here, we report millimeter wavelength observations of GN-z11 using the Northern Extended Millimeter Array (NOEMA), targeting the dust continuum and the \Cii$\,158\,{\rm \mu m}$ emission line.
Using the highly sensitive data obtained for GN-z11, we constrain its ISM properties, dust-obscuration, as well as dust production efficiency by comparing with dust enrichment and geometry models from \citet{Ferrara2022}.

This paper is organized as follows: in \S\ref{sec:observation}, we summarize the NOEMA observations so far performed for GN-z11. In \S\ref{sec:anallysis}, we present our data analysis and measurements for the \Cii$\,158\,{\rm \mu m}$ emission line and $\lambda\sim160\,{\rm \mu m}$ dust continuum. In \S\ref{sec:results}, we present the results. We present discussions in \S\ref{sec:discussion}. Finally, we conclude with the summary in \S\ref{sec:conclusion}. Throughout this paper, we assume a cosmology with $(\Omega_m,\Omega_{\Lambda},h)=(0.3,0.7,0.7)$, and the Chabrier \citep{Chabrier2003} initial mass function (IMF), where applicable.
With these cosmological parameters, 1 arcsec corresponds to $4.0\,{\rm kpc}$ in proper coordinate and the luminosity distance is $1.11\times10^5\,{\rm Mpc}$  at $z=10.603$.

\section{Observations} \label{sec:observation}

\subsection{NOEMA Observations}

Over the past years, NOEMA has targeted GN-z11 to search for the \Cii$\,158\,{\rm \mu m}$ emission line based on the redshift estimated from its early photometric and grism data from HST to explore its emission in the FIR. Deep NOEMA observations were conducted as separate observation projects across four cycles in the years 2014, 2016, 2018, and 2019. We summarize these observations in the following:

\vspace{0.1cm}
\noindent{\it Observations in 2014}: Using the WideX correlator, GN-z11 was observed by scanning the frequency range between 161.9 and 176.3 GHz (PI: C. Carilli). With 6 antennae, 4 different frequency settings were used. Each tuning has a 3.6GHz bandwidth.
The total on-source time was 46.6 hours, most of the data were taken in good to excellent conditions, and little flagging was required.
The spectral scan observations aimed to search for [CII] $158\,\mathrm{\mu m}$ emission line covering redshifts of $z\sim 9.78 - 10.74$, i.e., a photometric redshift range estimated based on the HST and Spitzer photometry \citep{Bouwens2015, Oesch2015}.

\vspace{0.2cm}
\noindent{\it Observations in 2016}: Using a single tuning with the WideX correlator, GN-z11 was observed with 8 antennae (PI: F. Walter and P. Oesch).
The on-source time was 5.2 hours in good weather conditions. Data calibration was done in a standard manner at IRAM, and only minor flagging was required.
The frequency between $155.2 - 158.8\,\rm{GHz}$ was tuned to target [CII] emission line at redshifts between $10.97-11.25$ based on the updated redshift estimation using the HST grism spectroscopy \citep{Oesch2016}.

\vspace{0.2cm}
\noindent{\it Observations in 2018}: Deeper spectral scans were performed for GN-z11 with 9 antennae and with the newly installed Polyfix correlator in 2018 (PI: F. Walter and P. Oesch). With a total on-source time of 17.4 hours, two Polyfix tunings were used to cover the frequency range between $152.9-163.6\,\rm{GHz}$.
These observations aimed to observe \Cii\ within the redshift range of $z_{\rm [CII]}=10.62-11.43$. Compared to the 2016 observations, these observations were deeper and covered a wider frequency range. 
The weather conditions were good. Data calibration was performed in a standard manner at IRAM, and minor flagging was required.

\vspace{0.2cm}
\noindent{\it Observations in 2019}: Follow-up observations of GN-z11 were again performed with 10 antennae with the Polyfix correlator (PI: F. Walter and P. Oesch). Total on-source time was 11.4 hours. A single tuning of the Polyfix correlator covered $156.9-164.6\,\rm{GHz}$, covering \Cii\ redshifts between $z_{\rm [CII]}=10.55 - 11.11$. Weather conditions were generally good and minor flagging was required.
We note that the observations in 2019 eventually covered the \Cii\ emission line from the actual spectroscopic redshift \citep[$z_{\rm spec}=10.60$;][]{Bunker2023} of GN-z11 using the updated NOEMA interferometer.

All the calibrations of phase, absolute flux, and amplitude were performed using the GILDAS software\footnote{\url{https://www.iram.fr/IRAMFR/GILDAS/}} with the support of IRAM astronomers.

\subsection{JWST Observations}

We obtained near-IR (NIR) images of GN-z11 from a JWST cycle-1 medium program entitled ``First Reionization Epoch Spectroscopic Complete Observation'' (FRESCO; \citealt{Oesch2023}). We used the F444W image as a sky location prior for extracting a spectrum and searching for dust continuum emission associated with GN-z11.
A comprehensive description of the survey design, data reduction, and analysis can be found in \citet{Oesch2023}.

Using JWST NIRCam \citep{RiekeNircam}, FRESCO performed deep slitless grism observations in the northern and southern GOODS/CANDELS fields using F444W filters with a medium spectral resolution of $R=1600$.
Simultaneously, broad- and medium-band images of the fields were acquired using the F444W, F210M, and F182M bands. The imaging integration times were $0.26\,\rm{hours}$, $1\,\rm{hours}$, and $1.2\,\rm{hours}$ per pointing for F444W, F210M, and F182M bands, respectively. 
The data were reduced using the \texttt{grizli} software (v1.7.11), which is publicly available \citep{grizli,grizli2}. The $5\,\sigma$ depth of the observations for the F444W imaging is $\sim28.2\,{\rm mag}$ for imaging observations \citep{Oesch2023}.



\section{Analysis} \label{sec:anallysis}

\subsection{Dust Continuum, IR Luminosity, and Dust Mass}

\begin{figure}
    \centering
    \includegraphics[width=1\columnwidth]{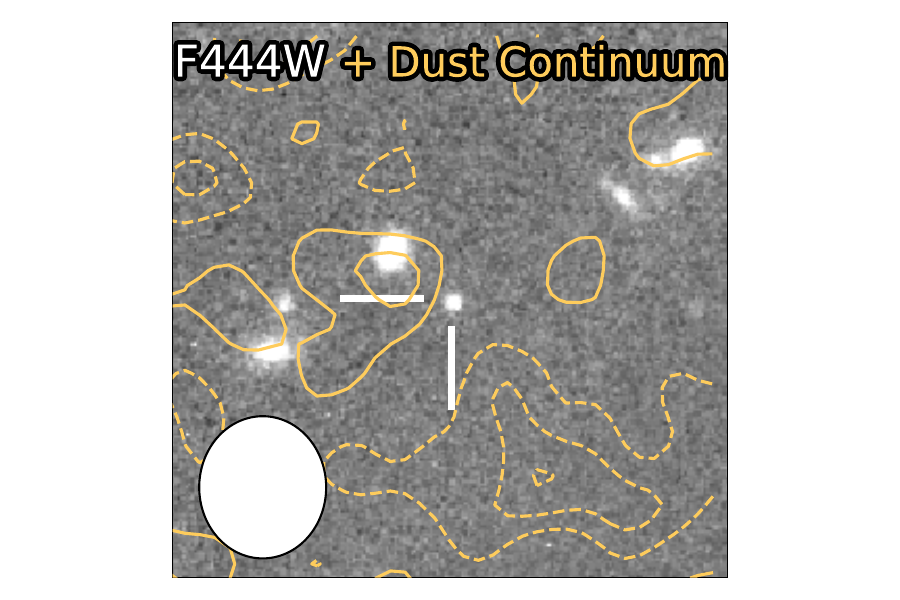}
    \caption{$8^{\prime\prime}\times8^{\prime\prime}$ cutout of the JWST F444W image obtained by the FRESCO JWST survey (background; \citealt{Oesch2023}) and dust continuum (orange contours) of GN-z11: Solid contours show 1, 2$\,\sigma$ and dashed contours show -3, -2, -1$\sigma$ where $1\,\sigma=13.0\,\mathrm{\mu Jy/beam}$. The white ellipse in the lower left corner shows the synthesized beam FWHM of the combined dust continuum image ($2.1^{\prime\prime}\times1.8^{\prime\prime}$).
    }
    \label{fig:continuum}
\end{figure}

Using the GILDAS package \texttt{CLIC}, the combined continuum uv-table of GN-z11 was made using all existing NOEMA data.
We then imaged the continuum data using the GILDAS \texttt{Mapping} package using a pixel size of $0.4^{\prime\prime}$. The resulting continuum map has a synthesized beam full-width-at-half-maximum (FWHM) of $2.05^{\prime\prime}\times1.83^{\prime\prime}$, and an RMS of $13\,\rm{\mu Jy/beam}$.
In the continuum image, we only find a $\sim2\,\sigma$ signal $\sim1^{\prime\prime}$ offset from GN-z11, and found no clear signals co-located with the JWST NIRCam F444W detection of GN-z11 (Fig. \ref{fig:continuum}). Thus, we concluded that the dust continuum of GN-z11 is not detected from the existing NOEMA observations.

As the dust continuum is not detected, we estimated upper limits for the dust continuum emission and IR luminosity. Using the
RMS of the continuum image, we obtained a $3\,\sigma$ upper limit of $<39\,\rm{\mu Jy}$. The upper limit of IR luminosity is then estimated by assuming a FIR SED for the GN-z11 by integrating a modified blackbody function between the rest-frame wavelength of $8-1000\,{\rm \mu m}$.
In particular, we assumed a dust temperature of $T_{d}=82\,\rm{\rm K}$ by extrapolating a dust temperature evolution estimated in \citet{Sommovigo2022}\footnote{We assumed transmissivity of the ISM to be $T=0.9$ as assumed by \citet{Fujimoto2022}, and a gas phase metallicity of $0.04\,{\rm Z_{\odot}}$ obtained in the SED fitting in \citet{Tacchella2023}} .
The derived dust temperature of $82\,\rm{\rm K}$ is much higher (by $\gtrsim30\,\rm{K}$) than those of $z\sim7$ galaxies \citep[e.g.,][]{Algera2023} but consistent with that currently estimated from one of the highest redshift dust detection at $z=8.31$ \citep[$>80\,{\rm K}$;][]{Bakx2020}. We discuss impacts of assuming a lower dust temperature in \S\ref{sec:tdunc}.
We also assumed a dust emissivity index of $\beta=2.0$ as is also assumed when estimating $T_{\rm d}$ in \citet{Sommovigo2022}, and applied a correction of the cosmic microwave background (CMB) attenuation based on \citet{dacunha2013}.
Due to the high CMB temperature of $31.7\,\rm{K}$ at $z=10.603$, the CMB attenuation with the assumed dust temperature is large and we need to multiply  the dust continuum emission by $1/0.88$.
Finally, we obtained a $3\,\sigma$ upper limit for the IR luminosity of $<3.6\times10^{12}\,{\rm L_{\odot}}$.


Using the $3\,\sigma$ continuum upper limit and the assumed dust temperature of $T_{\rm d}=82\,{\rm K}$, we estimated a dust mass upper limit in GN-z11. We used a dust mass absorption coefficient of $\kappa_{d}(\lambda_{\mathrm{rest}})=30\times(100\,{\rm \mu m}/\lambda_{\mathrm{rest}})^{\beta}\,\mathrm{cm^{2}\,g^{-1}}$ \citep{Inoue2020}, which is estimated as a typical value from different dust compositions and temperatures using laboratory experiments \citep[e.g.,][]{Mennella1998, Chihara2001,Boudet2005}. We then applied a correction for the CMB 
and obtained a dust mass upper limit of ${\rm log\,}(M_{\rm d}\,/\,\rm{M_{\odot}}) < 6.5$ ($3\,\sigma$).

\subsubsection{Dust Temperature Uncertainty} \label{sec:tdunc}
The typical dust temperature of star-forming galaxies at very high redshift is unconstrained from observations. Moreover, the assumed dust temperature has a large impact on the estimated IR luminosity (see discussion e.g., by \citealt{Ferrara2017,Ma2019,Fudamoto2020,Faisst2020,Lower2023}). 
Especially, our assumed dust temperature is much higher than those directly measured in lower redshift galaxies (e.g., $T_{\rm d}\sim45\,{\rm K}$ at $z\sim5-7$; \citealt{Liang2019,Bethermin2020,Faisst2020, Algera2023}).
Thus, to test the assumed dust temperature and to understand its impact on the IR luminosity upper limit, we employed an alternative method to estimate a dust temperature limit.

We tested the assumed dust temperature using the \texttt{FIS22} code presented in \citet{Fudamoto2022}. \texttt{FIS22} uses simple assumptions about star-to-ISM geometry and the radiation equilibrium between dust attenuation and re-emission. Using the measured size of the star-forming region, the code allows us to obtain estimations of dust temperature from a single FIR continuum observation.
To estimate the limit, we assumed the size of the star-forming region from the UV emission (i.e., $R_{\rm e}=64\pm20\,{\rm kpc}$; \citealt{Tacchella2023}), and the average ISM clumpiness parameter of ${\rm log\,\xi_{\rm clp}}=-1.02$ as found in \citet{Fudamoto2022}.
With a fixed size and an upper limit on FIR emission, we are able to obtain a lower limit on dust temperature. With the assumed dust size and UV luminosity of $L_{\rm UV}=7.40\pm0.01\times10^{11}\,\rm{L_{\odot}}$, we obtained $T_{\rm d} > 54\,{\rm K}$.
The derived lower limit indeed supports the high dust temperature of $T_{\rm d}$ assumed in our calculation. We note that the estimated dust temperature and its limit is still uncertain as the method assumes that the dust has the same spatial distribution as the UV emission. Nevertheless, we use the $T_{\rm d} = 54\,{\rm K}$ as a lower limit case as it currently gives the lowest $T_{\rm d}$ estimation.

In this lower limiting dust temperature case of $T_{\rm d}=54\,{\rm K}$, we find  the $3\,\sigma$ upper limit of IR luminosity to be $<7.7\times10^{11}\,{\rm L_{\odot}}$ (i.e., $\sim\times5$ smaller than that found by assuming $T_{\rm d}=82\,{\rm K}$). We further noticed that a dust temperature assumption lower than $50\,{\rm K}$ does not change the IR luminosity upper limit as the CMB correction becomes large and mitigates the difference of the dust temperature assumption. Thus, a systemic uncertainty of a factor $\sim5$ is very conservative. In the following discussions, we derive several estimates based on assuming the $82\,{\rm K}$ and $54\,{\rm K}$ dust temperatures.

\subsection{\Cii$\,158\,{\rm \mu m}$ Emission Line}

\begin{figure*}
    \centering
    \includegraphics[width=0.95\textwidth]{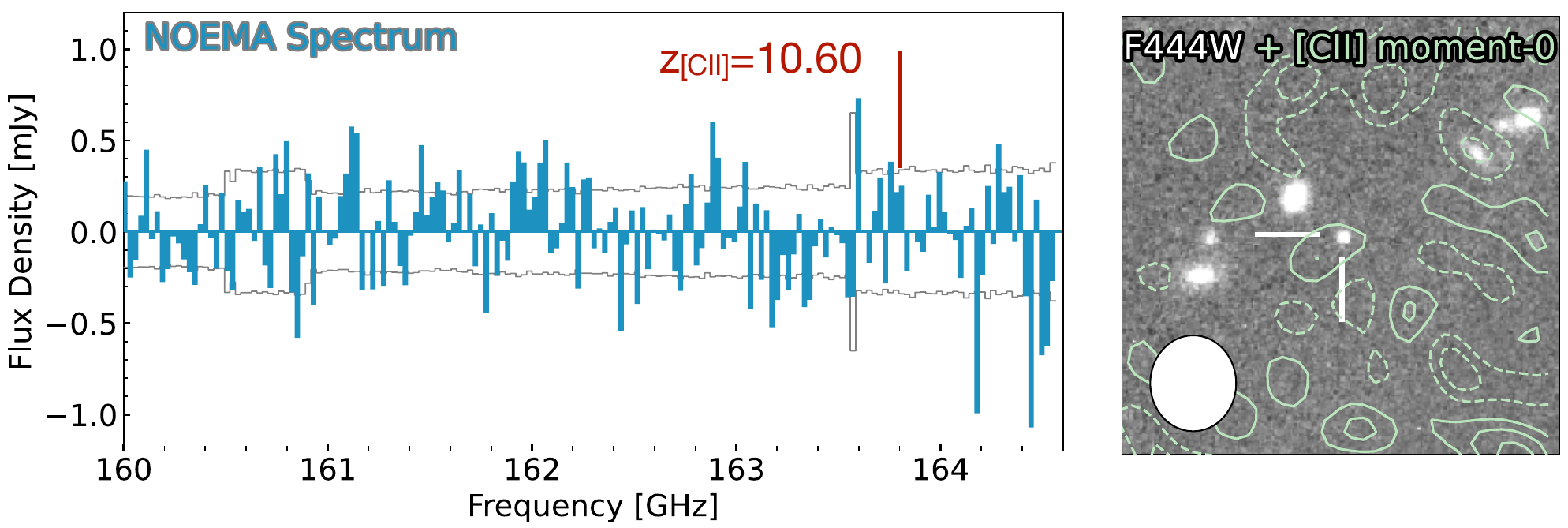}
    \caption{
    {\bf Left Panel: } The NOEMA spectrum of GN-z11 with $50\,\mathrm{km/s}$ binning. The gray solid line shows the RMS of each channel. The NOEMA observation covers \Cii$\,158\,\rm{\mu m}$ emission line of GN-z11 at the observed frequency of $\nu_{\rm obs}=163.84\,{\rm GHz}$ with the $z_{\rm spec}=10.60$ (Red line). From the data cube and extracted spectrum, we do not find any signal of the \Cii\ emission line.
    {\bf Right panel:} $8^{\prime\prime}\times8^{\prime\prime}$ cutout of JWST F444W image (background) and \Cii\ emission line moment-0 map (contours) of GN-z11. Contours show 1, 2, 3$\sigma$ and dashed contours show -3, -2, -1$\sigma$. The moment-0 map of the data cube is made by integrating over the $150\,\rm{km/s}$ of the \Cii\ emission line frequency. RMS of the image is $29\,{\rm mJy\,km/s}$, providing a $3\,\sigma$ upper limit of \Cii\ luminosity $<1.7\times10^{8}\,{\rm L_{\odot}}$.
    }
    \label{fig:c2}
\end{figure*}

The sky frequency of the \Cii$\,158\,{\rm \mu m}$ emission line of GN-z11, $\nu_{\rm [CII]-sky} = 163.791\,{\rm GHz}$, is covered by the observations in 2014 and 2019. From these observations, we focus on the observations of 2019 in our analysis which has much higher sensitivity than that of 2014 by using upgraded receivers with ten antennae. Using the \texttt{GILDAS} \texttt{Mapping} package, we created a data cube with $50\,{\rm km/s}$ spectral binning with a pixel angular size of $0.4^{\prime\prime}$. The resulting data cube has a RMS of $\sim 0.31 \,{\rm mJy/beam}$ in the $50\,{\rm km/s}$ binning at $\nu = 163.791\,{\rm GHz}$.

We then extracted a spectrum of GN-z11 using a circular aperture with a radius of $r=1^{\prime\prime}$ on top of the JWST detection of GN-z11 (left panel of Fig. \ref{fig:c2}). The $1^{\prime\prime}$ radius is consistent with the NOEMA beam size. At the frequency of \Cii\ at $z=10.603$, we only find $< 1\,\sigma$ signals in the spectra. To search for any spatially off-set signals, we further created a \Cii$\,158\,\rm{\mu m}$ emission moment-0 map. To create the \Cii\ moment-0 image, we assumed the line width of \Cii\ emission to be $150\,{\rm km/s}$, which is consistent with recent FIR emission line observations of $z>9$ galaxies \citep[e.g.,][]{Hashimoto2018}.
In the surrounding area of the GN-z11, we only find $\sim1\,\sigma$ signals (Fig. \ref{fig:c2}) which is much lower than the typical secure detection thresholds for lines with well-known sky location and spectroscopic redshift (e.g., $>3.5\,\sigma$; \citealt{Bethermin2020}).
Based on these results, we concluded that the \Cii$\,158\,\rm{\mu m}$ emission line is not detected in the current NOEMA data. 

Based on the non-detection, we measured a $3\,\sigma$ upper limit of the \Cii\ emission line luminosity by measuring three times the pixel-by-pixel RMS of the moment-0 map. 
We obtained a $3\,\sigma$ \Cii\ flux upper limit of $0.087\,\rm{Jy\,km/s}$, and  \lcii\ upper limit of GN-z11 of $<1.7\times10^{8}\,{\rm L_{\odot}}$.

\renewcommand{\arraystretch}{1.3}
\begin{table}[]
    \centering
    \begin{tabular}{ll}
    \hline
    \multicolumn{2}{c}{NOEMA Measurements} \\
    \hline
    $f_{\rm [CII]}$ & $<29\,{\rm mJy\,km/s}$ ($3\,\sigma$ limit) \\
    ${\rm log\,}(L_{\rm [CII]}\,/\,\rm{L_{\odot}})$  & $<8.2$ ($3\,\sigma$ limit) \\  
    $f_{\rm cont, 160\,{\mu m}}$ & $<39\,\rm{\mu Jy}$ ($3\,\sigma$ limit) \\
    ${\rm log\,}(L_{\rm IR}\,/\,\rm{L_{\odot}})$  & $<12.5$ (with $T_{\rm d}=82\,{\rm K}$)\\
    & $<11.9$ (with $T_{\rm d}=54\,{\rm K}$) \\
    ${\rm log\,}(M_{\rm d}\,/\,\rm{M_{\odot}})$ & $< 6.5$ (with $T_{\rm d}=82\,{\rm K}$)\\
    & $< 6.9$ (with $T_{\rm d}=54\,{\rm K}$) \\
    & and $\kappa_{d}(\lambda_{\mathrm{rest}})=30\times(100\,{\rm \mu m}/\lambda_{\mathrm{rest}})^{\beta}\,\mathrm{cm^{2}\,g^{-1}}$) \\
    \hline
    \end{tabular}
    \caption{Measured properties of GN-z11 from NOEMA observations.}
    \label{tab:measurements}
\end{table}

\section{Results} \label{sec:results}

\subsection{Dust Obscured Star Formation Activity of GN-z11}

With the relatively high upper limit of \lir, the dust-obscured star formation rate is only weakly constrained with ${\rm SFR_{IR}} < 430\,{\rm M_{\odot}\,yr^{-1}}$ with $T_{\rm d}=82\,{\rm K}$ and $<92\,{\rm M_{\odot}\,yr^{-1}}$ with $T_{\rm d}=54\,{\rm K}$ assuming a conversion of ${\rm SFR_{IR}}=1.2\times10^{-10}\,L_{\rm IR}\,{\rm M_{\odot}\,yr^{-1}\,L_{\odot}^{-1}}$ \citep{Madau2014,Inami2022}. The weak upper limit on ${\rm SFR_{IR}}$ despite the deep continuum observations is due to the assumed high dust temperature of $T_{\rm d}=82\,{\rm K}$ and $54\,{\rm K}$. Using these limits, the infrared excess (IRX$={\rm log}(L_{\rm UV}/L_{\rm IR})$; \citealt{Meurer1999}) $3\,\sigma$ limit of GN-z11 is estimated to be $>-0.9$ to $>-1.5$. To further constrain the dust obscured star formation activity of GN-z11, accurate constraints on the dust temperature is essential via deep observations at wavelengths $<1\,{\rm mm}$.

With the estimated stellar mass of $\sim10^{9}\,{\rm M_{\odot}}$ \citep{Tacchella2023}, the dust mass upper limit shows a stellar-to-dust mass ratio of $\xi_{\rm d}\lesssim 0.003$ for $T_{\rm d}=82\,{\rm K}$ and $\xi_{\rm d}\lesssim 0.008$ for $T_{\rm d}=54\,{\rm K}$. These limits are consistent with typical values in lower redshift galaxies \citep[e.g.,][]{Calura2017,Dayal2022}. 

\subsection{\Cii$\,158\,\rm{\mu m}$ Emission Line}

\begin{figure}
    \centering
    \includegraphics[width=0.95\columnwidth]{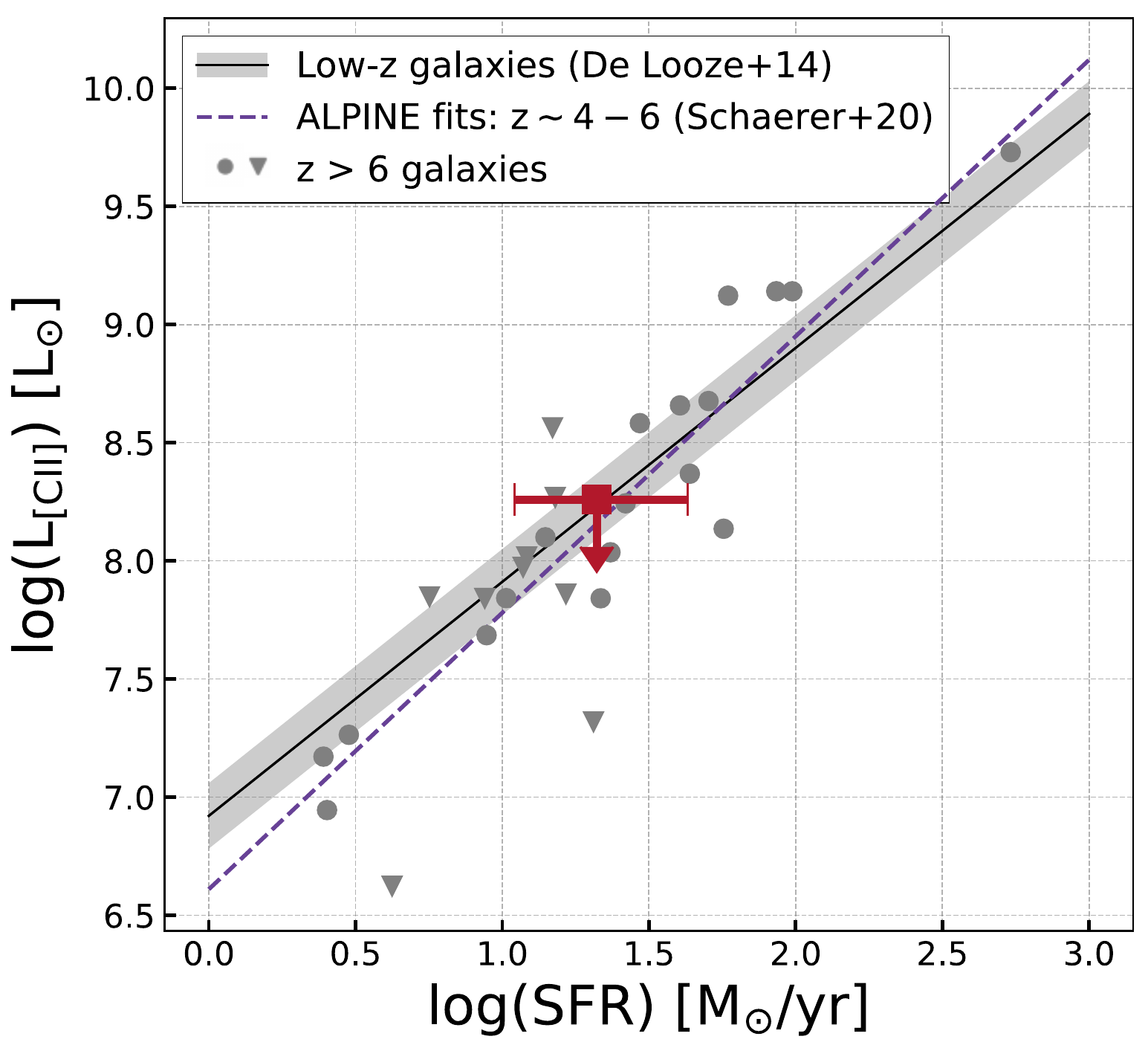}
    \caption{Star formation rate versus \Cii\ emission line luminosity of GN-z11. Previous observations of $z>6$ are also plotted with gray points (\citealt{Harikane2020}; \citealt{Fudamoto2023}; \citealt{Schouws2022} and references therein). Downward triangles show $3\,\sigma$ upper limit in case of non-detections. Lines show previously obtained relations for low- and high-redshift galaxies (solid: \citealt{Schaerer2020}, Dashed: \citealt{Delooze2014}).
    For the SFR of GN-z11, we adopted the SED fitting results of $21^{+22}_{-10}\,{\rm M_{\odot}\,yr^{-1}}$ from \citet{Tacchella2023}.}
    \label{fig:C2SFR}
\end{figure}

To compare the star formation activity and \lcii\ of GN-z11, we adopted the total star-formation rate (SFR) of $21^{+22}_{-10}\,{\rm M_{\odot}\,yr^{-1}}$ obtained in the spectral energy distribution (SED) fitting in \citet{Tacchella2023}. We find that based on the upper limit \lcii\ from our measurement, GN-z11 has \lcii\ consistent with or slightly smaller than that of low redshift similarly star-forming galaxies (Fig. \ref{fig:C2SFR}). Although few observational studies of the SFR-\lcii\ relation for galaxies at $z>10$, this shows that the \Cii\ luminosity of GN-z11 is not enhanced compared with the relation obtained for star-forming galaxies at lower redshifts. 

The existence of an active galactic nucleus (AGN) was proposed in GN-z11 based on the significant detection of [Ne{\sc iv}]$\,\lambda 2423$ using deep NIRSpec observation \citep{Maiolino2023a}. If the rest-frame UV emission of GN-z11 is dominated by AGN activity, the SFR of GN-z11 might be overestimated and thus our constraints on the SFR-$L_{\rm [CII]}$ relation of GN-z11 should be lifted.

\section{Discussion}
\label{sec:discussion}

\subsection{Molecular gas mass and gas depletion time scale at $z=10.6$}

\begin{figure}
    \centering
    \includegraphics[width=0.99\columnwidth]{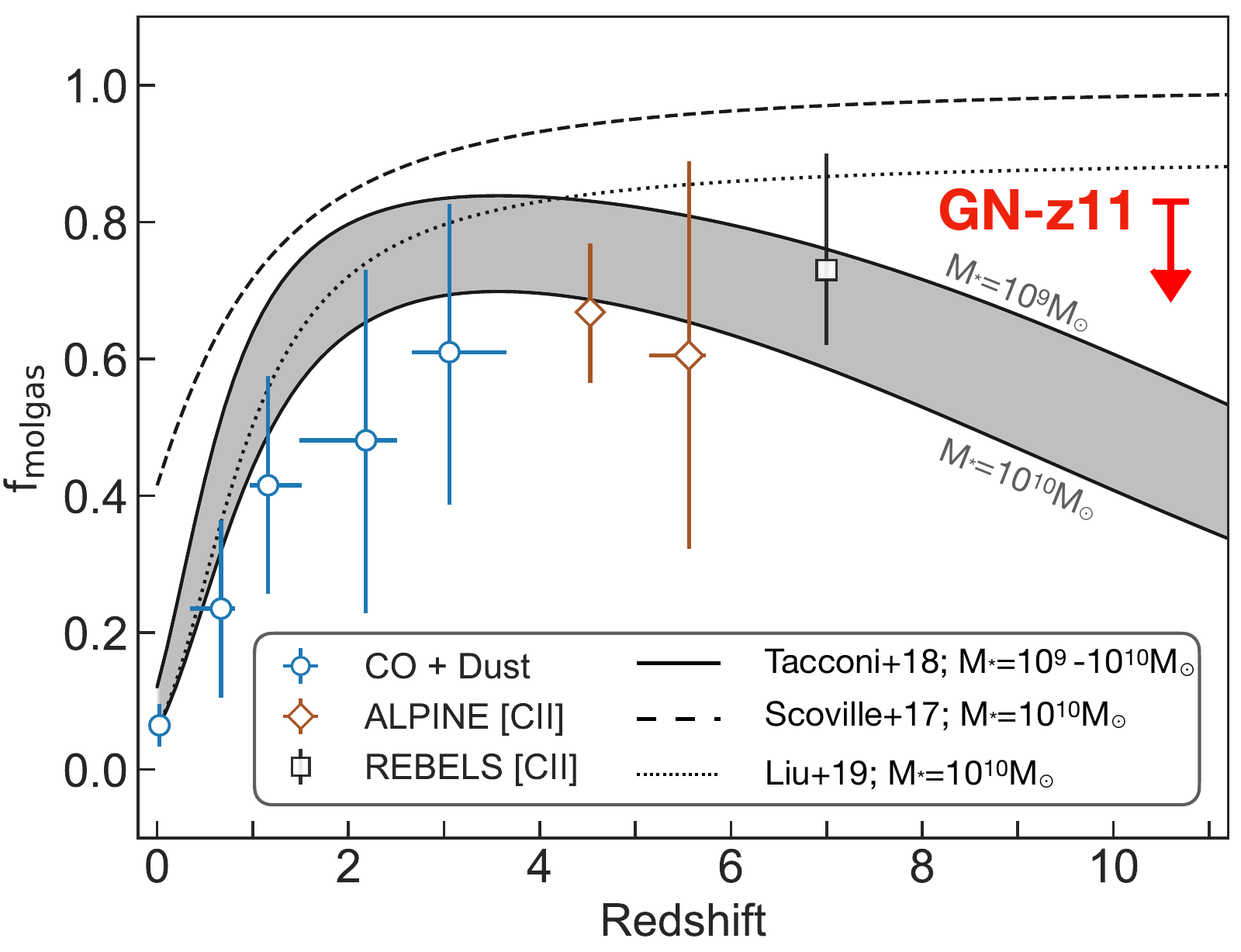}
    \caption{Evolution of the molecular mass fraction ($f_{\rm molgas} = M_{\rm molgas}/(M_{\rm molgas} + M_{\rm star0 })$) across a wide redshift range.
    Various indicators are used to estimate molecular gas mass: CO emission lines and dust continuum at $z<4$ (\citealt{Dessauges2017,Scoville2017,Tacconi2018,Tacconi2020}, see also \citealt{Boogaard2020}), \Cii\ emission lines at $4<z<6$ from the ALPINE survey \citep{Dessauges2020} and at $z\sim7$ from the REBELS survey (Aravena et al. submitted). Data points for the $z<8$ samples represent mean values and errorbars are 16th to 84th percentiles within the distribution. Upper limit of GN-z11 is within the range of $z>4$ galaxy observations.
    }
    \label{fig:fmol}
\end{figure}

Recent studies showed that the dominant fraction of \Cii$\,158\,{\rm \mu m}$ emission lines arise from neutral gas, such as neutral photo-dissociation regions (PDRs). In particular, luminous galaxies typically have $>80\%$ of the \Cii\ emission originate from the PDRs \citep{DiazSantos2017}. For high-redshift galaxies, this feature is further supported by detailed calculations and observations of emission lines arising from HII regions \citep{Decarli2014,Witstok2022,Decarli2022}. This makes the \Cii$\,158\,{\rm \mu m}$ emission line an excellent tracer of star-forming neutral gas \citep[e.g.,][]{Zanella2018,Sommovigo2021}.


Using the formalism presented in \citet{Zanella2018}, we estimate the upper limit of the molecular gas mass of GN-z11 to be ${\rm log\,}(M_{\rm mol}\,/\,\rm{M_{\odot}}) < 9.7$. With the estimated stellar mass of $\sim10^{9}\,{\rm M_{\odot}}$ \citep{Tacchella2023}, the molecular gas fraction ($f_{\rm mol} = M_{\rm mol}/(M_{\rm star} + M_{\rm mol})$) of GN-z11 is $f_{\rm mol}<0.83$ ($3\,\sigma$).  The upper limit from the \Cii\ observations is largely consistent with lower redshift galaxies (Fig. \ref{fig:fmol}). With the estimated star formation rate of $21\,{\rm M_{\odot}\,yr^{-1}}$ \citep{Tacchella2023}, the molecular gas depletion time of GN-z11 is $t_{\rm depl} < 0.2\,{\rm Gyr}$ ($3\,\sigma$). While the contribution from the AGN to the UV-luminosity is needed to estimate a more accurate SFR, the result suggests that GN-z11 might deplete all molecular gases before $z\gtrsim8$, unless molecular gas is rapidly replenished by accretion.

Recent CO and [CI] emission line observations using $z\sim6$ AGNs suggest that \Cii\ based molecular gas mass measurements could be systematically higher by $\sim0.5\,{\rm dex}$ compared with those estimated from [C{\sc i}] and/or CO observations \citep{Neeleman2021,Decarli2022}. Thus if the $\sim0.5\,{\rm dex}$ over-prediction is the case for our estimations, the upper limit might need to be systematically lowered, providing stronger constraints.
Deeper observations of the \Cii\ emission line and observations of other molecular gas indicator (e.g., emission lines from [CI] and/or CO) will be crucial to further constrain these essential parameters of galaxy growth.

\subsection{Expected dust opacity and constraints on the dust formation}
\label{sec:Ferrara}
Here, we compare our dust continuum upper limits with recent theoretical analysis of $z>10$ UV-luminous galaxies \citep[e.g.,][]{Ferrara2023}.

\citet{Ferrara2023} discussed that the JWST based finding of very blue and compact galaxies was surprising. 
Indeed, the observed dust optical depth at the rest-frame $1500\,{\rm \AA}$ deduced for GN-z11 is $\tau_{1500}=0.49$ (corresponding to $A_V=0.17$, \citealt{Bunker2023b}). This low value does not agree with the expected dust optical depth estimated from the extremely compact effective radius ($r_{\rm e} = 64\,{\rm pc}$; \citealt{Tacchella2023}) and the dust mass associated with the estimated massive stellar mass ($\sim10^{9}\,{\rm M_{\odot}}$). By assuming a standard dust-to-stellar mass ratio $\xi_d\sim0.002$, appropriate for a Salpeter IMF ($1-100\,{\rm M_{\odot}}$) and a dust yield of $0.1\,{M_{\odot}}$ per supernova, we estimate a dust mass of $M_{\rm d}\sim10^6\,{\rm M_{\odot}}$ \citep[e.g.,][]{Hirashita2014}. If UV and dust are co-located in the tiny $\sim64\,{\rm pc}$ region, using the calculation performed in \citet{Ferrara2023}, we estimated that the resulting optical depth would be $\tau_{1500} \sim 500$, i.e. far exceeding the observed value.

The dust envelope can be made less opaque if the dust has a spatial distribution that is much more extended than the stellar $r_e$. In particular, assuming a spherical dust distribution, the observed value $\tau_{1500} = 0.49$ is recovered if the dust distribution is expanded out to $r\sim30\times r_{\rm e}$ to  $100\times r_{\rm e}$ (i.e., $\sim2\,{\rm kpc}$ to $\sim6.4\,{\rm kpc}$). Physically, this configuration can be achieved, for example, as a result of an outflow emanating from the galaxy. The presence of radiation-driven dusty outflows could arise from the super-Eddington nature of galaxies like GN-z11 \citep{Ferrara2023, Ziparo2023, Fiore2023}.


In our observations, we find a $\lambda_{\rm rest}=160\,{\rm \mu m}$ dust continuum of $< 13\,{\rm \mu Jy}$ at $1\,\sigma$ ($<26\,{\rm \mu Jy}$, $2\,\sigma$). This value is higher than the predicted one at any stage during the outflow expansion. In fact, when the outflow has expanded the dust distribution so as to produce the observed value of $\tau_{1500}$, we predict that the $160\,{\rm \mu m}$ flux should be $4.5\,{\rm \mu Jy}$, hence about 3 times below our upper limit. Clearly, uncertainties are related to the assumed value of $\xi_d$. The present observations, however, can be used to obtain a limit on $\xi_d < 0.014$ at $1\,\sigma$ ($\xi_d < 0.32$ at $2\,\sigma$), which translates a predicted upper limit on the dust mass of $<7\times 10^6\,{\rm M_{\odot}}$ $(<1.6\times 10^8\,{\rm M_{\odot})}$. Thus, at least $3\times$ deeper continuum observations would be crucial to provide constraints on dust formation at $z>10$  and to test the proposed scenario to explain the very blue color of high-redshift massive compact galaxies.

\section{Conclusion} \label{sec:conclusion}
In this paper, we present results based on NOEMA observations of GN-z11 targeting the \Cii$\,158\,\rm{\mu m}$ emission line and the underlying dust continuum from a UV luminous star-forming galaxy GN-z11 at $z=10.60$. We found the following results: 

\vspace{0.15cm}
\noindent$\bullet$ Although NOEMA observations were extremely deep and covered the sky frequency of the \Cii$\,158\,{\rm \mu m}$ emission line,
 we did not detect either \Cii$\,158\,{\rm \mu m}$ or dust continuum. We reported $3\,\sigma$ upper limits of the \Cii\ luminosity ($L_{\rm [CII]]} < 1.7\times10^8\,{\rm L_{\odot}}$) and the IR luminosity ($L_{\rm IR} < 3.6\times10^{12}\,{\rm L_{\odot}}$).

\vspace{0.15cm}
\noindent$\bullet$ Based on the upper limits, we found that GN-z11 is not bright either in \Cii\ emission or dust continuum.
The upper limit \lcii\ of GN-z11 is consistent or could be slightly smaller than that predicted from the SFR-\lcii\ relation observed from lower redshift galaxies at $z<8$.

\vspace{0.15cm}
\noindent$\bullet$ As \Cii\ emission mostly traces the neutral ISM, we estimate a molecular gas mass for GN-z11 using the relation in \citet{Zanella2018}. We find an upper limit of $M_{\rm mol} < 5.0\times10^{9}\,{\rm M_{\odot}}$, and a depletion time of $<0.2\,{\rm Gyr}$, suggesting that the GN-z11 could deplete all the star-forming gas by $z\gtrsim8$.

\vspace{0.15cm}
\noindent$\bullet$ The blue color ($\beta=-2.4$) suggests that GN-z11 is dust poor, which is consistent with the non-detection of dust continuum.
The current upper limit is $>\times 2$ above the expected continuum flux density expected assuming typical dust production.

Our observations of GN-z11 showcase the crucial constraints that submm/mm observations can provide for very early galaxies.

\section*{Acknowledgements}
This work is based on observations carried out under project number XACE, XBCE, XCCE, XDCE, S16CQ, W17FD, and W18FD with the IRAM NOEMA Interferometer. IRAM is supported by INSU/CNRS (France), MPG (Germany) and IGN (Spain).
YF, AKI, YS acknowledge support from NAOJ ALMA Scientific Research Grant number 2020-16B. YF further acknowledges support from support from JSPS KAKENHI Grant Number JP23K13149. PAO acknowledges support from the Swiss National Science Foundation through project grant 200020\_207349. The Cosmic Dawn Center (DAWN) is funded by the Danish National Research Foundation under grant No. 140. 
HD acknowledges financial support from the Agencia Estatal de Investigaci{\'o}n del Ministerio de Ciencia e Innovaci{\'o}n (AEI-MCINN) under grant (La evoluci{\'o}n de los c{\'i}umulos de galaxias desde el amanecer hasta el mediodía c{\'o}smico) with reference (PID2019-105776GB-I00/DOI:10.13039/501100011033).

\section*{Data Availability}

The data underlying this article will be shared on reasonable request to the corresponding author.



\bibliographystyle{mnras}
\bibliography{base} 








\bsp	
\label{lastpage}
\end{document}